# Attention-aware convolutional neural networks for identification of magnetic islands in the tearing mode on EAST tokamak

Feifei Long[1], Yian Zhao[1], Yunjiao Zhang[1], Chenguang Wan[2], Yinan Zhou[1], Ziwei Qiang[1], Kangning Yang[1], Jiuying Li[1], Tonghui Shi[2], Bihao Guo[2], Yang Zhang[2], Hailing Zhao[2], Ang Ti[2], Adi Liu[1], Chu Zhou[1], Jinlin Xie[1], Zixi Liu[1*], Ge Zhuang[1], EAST Team[2]

[1]University of Science and Technology of China, Hefei 230031, China

[2]Institute of Plasma Physics, Chinese Academic Sciences, Hefei 230031, China

E-mail: zxliu316@ustc.edu.cn

These authors contributed to the work equally and should be regarded as co-first authors.

**Abstract：**

  The tearing mode, a large-scale MHD instability in tokamak, typically disrupts the equilibrium magnetic surfaces, leads to the formation of magnetic islands, reduces core electron temperature and density, thus resulting in significant energy losses and may even cause discharge termination. This process is unacceptable for ITER. Therefore, the accurate identification of a magnetic island in real time is crucial for the effective control of the tearing mode in ITER in the future. In this study, based on the characteristics induced by tearing modes, an attention-aware convolutional neural network (AM-CNN) is proposed to identify the presence of magnetic islands in tearing mode discharge utilizing the data from ECE diagnostics in the EAST tokamak. A total of 11 ECE channels covering the range of core is used in the tearing mode dataset, which includes $2.5 \times 10^9$ data collected from 68 shots from 2016 to 2021 years. We split the dataset into training, validation, and test sets (66.5%, 5.7%, and 27.8%), respectively. An attention mechanism is designed to couple with the convolutional neural networks to improve the capability of feature extraction of signals. During the model training process, we utilized adaptive learning rate adjustment and early stopping mechanisms to optimize performance of AM-CNN. The model results show that a classification accuracy of 91.96% is achieved in tearing mode identification. Compared to CNN without AM, the attention-aware convolutional neural networks demonstrate great performance across accuracy, recall metrics, and F1 score. By leveraging the deep learning model, which incorporates a physical understanding of the tearing process to identify tearing mode behaviors, the combination of physical mechanisms and deep learning is emphasized, significantly laying an important foundation for the future intelligent control of tearing mode dynamics.

## 1. Introduction

  Tearing mode is a type of magnetohydrodynamic (MHD) instability that forms magnetic islands in tokamak plasmas, leading to enhanced plasma transport [1], core confinement reduction [2], and even plasma disruption [3]. The significant energy losses and disruption of equilibrium magnetic surfaces caused by tearing modes make their accurate detection crucial for the control of stable operation in fusion plasmas, particularly for ITER. There are many causes of disruptions, such as 2/1 neoclassical tearing mode [4], locking mode [5], density limit [6], specific pressure limit [7], core impurities accumulation [8], vertical instability [9], in which a 2/1 neoclassical tearing mode (NTM) [10] is treated as primary cause [11][12] due to the finite plasma resistivity of rational surface in the ITER baseline scenario [13]. Besides, NTMs are associated with large-scale magnetic islands that evolve slowly at rational surfaces with low ($m$, $n$) mode numbers in tokamaks [14]. As these magnetic islands grow and reach a critical size, they cause a decrease in the normalized beta [15], thereby limiting its further increase. This instability is primarily driven by the pressure-driven (bootstrap) current, which serves as a seed for the growth of NTM [16]. Therefore, controlling NTMs has become one of the primary challenges for achieving stable operation in future ITER.

Two main approaches are currently applied for controlling NTMs. First, during the early stage of seed island formation, accurate identification and characterization of the behavior of seed islands are focused on to gain deeper insight into their formation mechanisms, triggering conditions, and parameter space. Early intervention is then applied to prevent further development of the seed island. Secondly, once the seed island has formed and evolved into a saturated magnetic island, ECCD is used to drive currents that effectively limit further island growth. Both methods inevitably require the rapid identification of the formation and growth of magnetic islands. Due to traditional tearing mode recognition algorithms relying on specific feature extraction methods (such as fast fourier transform, wavelet transform, singular value decomposition) in complex plasma conditions in real-time applications, the manual parameters adjustment of these methods is required to meet the real-time performance and accuracy of mode recognition. Different from traditional algorithms, the formation process of plasma magnetic islands can be monitored and analyzed by the machine learning model, and various parameters can be dynamically adjusted according to real-time monitoring data to effectively control the onset of NTMs. This is because the deep learning algorithms allow models to adapt to subtle changes in signal characteristics and effectively identify complex tearing mode signals, which shows significant potential in tearing mode recognition and may play a crucial role in the precise identification and control of future seed magnetic islands.

In recent years, efforts have been made to employ traditional and intelligent methods for recognizing and controlling the tearing mode in tokamaks, and many results have been achieved. A traditional simulated framework on ITER [17] for evaluating neoclassical tearing mode detection using electron cyclotron emission (ECE) is developed, with advanced perturbation modeling introduced, ECE channel layouts optimized, and detection algorithms compared quantitatively to balance accuracy and latency for effective NTM control. In J-TEXT, a PID system for rapid island phase recognition has been developed to support tearing mode feedback [18]. In TCV tokamak, a transient analysis model based on the triggering characteristics of 2/1 NTMs has been proposed to evaluate the time-varying classical stability index during test discharges [19]. To achieve better detection and control of tearing modes, artificial intelligence techniques have been explored and applied. A real-time feedback control system based on machine learning algorithms was successfully developed and tested for avoiding tearing modes and disruptions in DIII-D plasmas, utilizing ensemble learning methods and real-time data from the plasma control system (PCS) to maximize plasma performance while ensuring stability [20]. A MHD pattern recognizer, combining a temporal convolutional network and a long short-term memory (LSTM) network, has been developed and tested on a dataset of 33 shots on EAST tokamak [21]. This recognizer can accurately calculate the frequency and intensity of the MHD mode, demonstrating the potential of deep learning for MHD mode identification on EAST [21]. Additionally, deep reinforcement learning has been successfully used to achieve normalized beta control in the KSTAR tokamak [22]. In a 2024 study of the DIII-D tokamak [23], it was demonstrated that deep reinforcement learning has been successfully applied to avoid tearing instabilities in fusion plasmas, proving its effectiveness in controlling plasma stability and operating performance.

In our work, we develop an attention-aware convolutional neural networks for automatic characteristic recognition of tearing mode using the muti-channels electron cyclotron emission (ECE) diagnostic on EAST tokamak. The signal intensities measured by local heterodyne radiometer are proportional to the cube of the electron temperature [24], making ECE system highly sensitive to onset of tearing instabilities. In addition, the ECE diagnostic has a temporal resolution of 10 μs and a radial resolution of approximately 1–2 cm [25], which is allowed to capture the temperature fluctuation caused by rotating magnetic islands. Hence, the multi-channels ECE diagnostic can be useful to identify islands in the tearing instability discharges. A dataset comprising $2.5 \times 10^9$ samples was collected from 68 discharges conducted between 2016 and 2021 to develop a classifier for identifying tearing modes on the EAST tokamak. We innovatively combined the attention mechanism with convolutional neural networks, grounded in physical principles, to accurately identify the tearing modes. The model's results demonstrate that the classifier can achieve a recognition accuracy of 91.96%. This approach enhances the interpretability, accuracy, generalization of the model by incorporating the physical mechanisms of the tearing mode. The organizational structure of this paper is outlined as follows. Section 2 provides a brief introduction to the basic structure and key parameters of the neural network. Subsection 2.1 present a detailed description of the neural network construction modules. Subsection 2.2 discuss the training data and processing methods in detail. Section 3 covers the training parameter settings and result analysis for the models discussed. Finally, the summary and discussion are presented in Section 4.

## 2. Database and Method

### 2.1 Dataset split and Experimental Setup

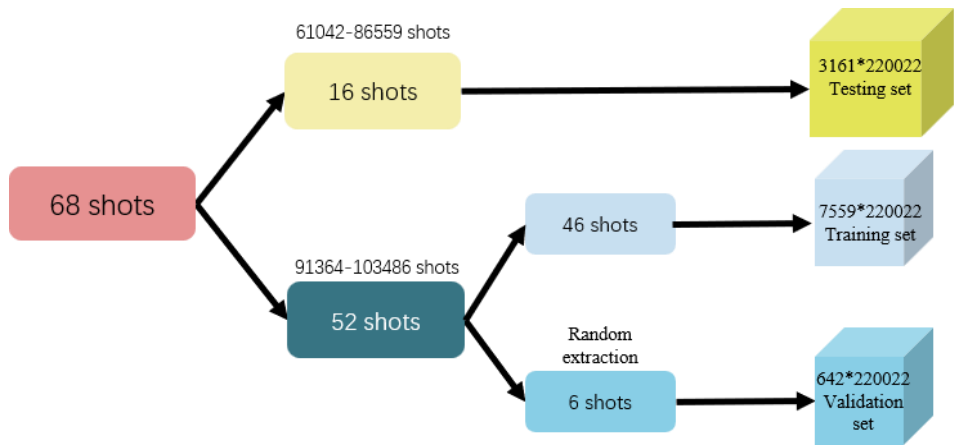

**Figure 1. Tearing mode dataset includes 11362×11×20002 data collected from 68 shots from 2016 to 2021 on the EAST tokamak. 16 shots selected from discharges #61042 to #86559. The remaining 52 shots, spanning from discharges #91364 to #103486, are further divided into the training and validation sets.**

Experimental signals of tearing mode on the EAST tokamak have been analyzed for the purpose of identifying the tearing instabilities. For tearing modes occurring on resonant surfaces in the core region, such as the (1,1) or (2,1) modes, localized measurements from diagnostics like ECE and SXR are used to detect the presence of magnetic islands. The dataset includes 68 discharges with tearing modes, spanning from shot 61042 (February 25, 2016) to shot 103486 (July 26, 2021). The dataset is split by discharges into training, testing, and validation sets. As shown in figure 1, 16 shots from shots 61042-86559 are designated as test sets, and 6 of the 52 shots from 91364-103486 are randomly chosen as validation sets for tuning parameters during training. The remaining 46 shots serve as the training set. The dataset consists of a total of 11362×220022 samples, with labels indicating the presence (label as 1) or absence (label as 0) of tearing modes. The labels are obtained from experimental diagnostic systems such as Mirnov probes, ECE systems and SXR diagnostic. The frequency spectrum of tearing mode is also confirmed by these diagnostics. The label presence of tearing modes is determined by performing a correlation analysis between the ECE signals and the magnetic probe signals. And the labels in the dataset correspond to the presence or absence of a tearing mode, and they have been manually curated and carefully selected for each individual shot. Each label is determined through a thorough inspection of the experimental data, ensuring a high level of accuracy.

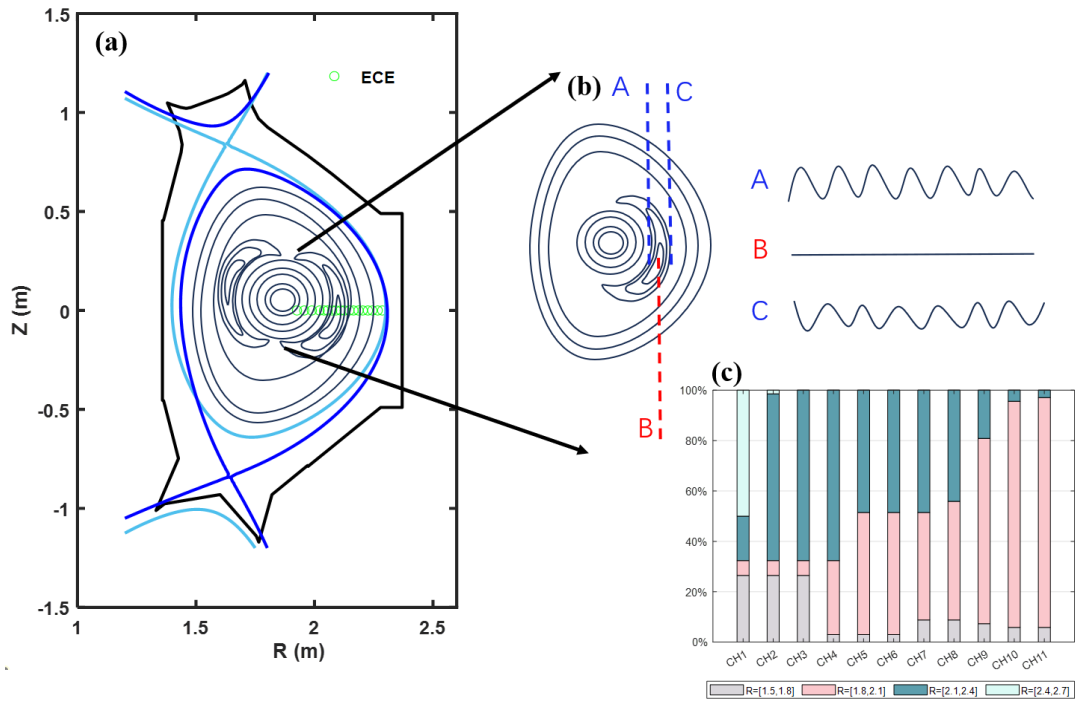

Figure 2: (a) A Diagrammatic of island structure in tearing instabilities shown in the EAST tokamak. (b) Time traces of ECE signals at different radial positions on the outside of island (point C) and inside of island (point A), the O point of island (point B). (c) Percentile pile bar chart of radial different positions in the databases from 11 channels of ECE. The major radius of EAST tokamak is 1.85 m.

The formation of tearing modes in a tokamak is typically preceded by certain physical phenomena, such as the onset of magnetic islands and sawtooth collapses. Consequently, the generation of magnetic islands is considered a key characteristic for the onset of tearing modes. As magnetic islands expand and rotate, as illustrated in figure 2(a), they often lead to dynamic changes in the temperature and density of the plasma core region. Due to the localized nature of ECE measurements and the minimal relative displacement, the rotastion of the magnetic island results in signal inversions at different positions of the island, such as the X-point and O-point. A phase difference resulted from island rotation can be utilized to detect the presence of a magnetic island and confirm the exist of a tearing mode. Figure 2(b) shows the time evolutions of electron temperature around an island located within the separatrix. Within the island, the pressure distribution flattens, while outside the islands it remains largely unchanged. When the signal passes through the magnetic island, the temperature profile deviates from a sinusoidal waveform. Additionally, it can be observed that the amplitude of temperature modulation increases from the O-point towards the separatrix of the magnetic island and then gradually decreases outside. A total of 11-channel signals with 68 shots has been selected as covering the radial region from the core to edge of plasma shown in figure 2 (c). Specifically, the original sampling rate of the ECE signal is 1 MHz, and the frequency of the tearing mode signal is usually between 1 kHz and 20 kHz. The ECE signals are randomly divided into time slices of varying lengths, ranging from 0.005 s to 0.02 s.

As shown in figure 2(c), the database in our work comes from the 1-11 channels data of the ECE signals, which can cover the region from the plasma boundary to near the plasma core, covering the area where tearing modes occur. After applying denoising process to the radiation signal intensity of the eleven channels, a feature vector of a sample connected afterwards is obtained. The feature vector of a sample connected afterwards is a sequence of length 220022 (11×20002). However, it is noted that the toroidal magnetic field of the discharges in the database ranges from 1.6 T to 2.2 T, and the positions of the electron cyclotron radiation signals detected by the ECE system are dependent on the magnetic field strength at different frequencies. Figure 2(c) presents a percentile histogram of the radial positions for the 11 signals in the 68 shots. As can be seen, the ECE signals from channels 1 to 11 are able to cover the radial region where the magnetic islands are present (approximately within 1.8 m < R < 2.1 m) in the dataset.

### 2.2 Model Architecture

In the following sections we describe the architecture of the attention aware convolutional neural network (named AM-CNN) and the algorithm of the attention mechanisms. Differing from traditional neural networks [26], the model incorporates an attention mechanism along with convolutional layers that automatically detect spatial hierarchies and local patterns in data. This makes it particularly effective for processing tearing mode feature, whereas standard convolutional networks typically rely solely on convolutional layers without attention mechanisms. To explore higher accuracy of intelligent recognition of tearing modes, we compare traditional neural networks with a custom-designed attention-aware CNN developed specifically for this study. This architecture integrates a dot-product attention with convolutional neural networks to better capture the features of reversal ECE signals at different positions. The innovation aims to overcome the limitations of traditional convolutional neural networks, which experience decreased recognition accuracy in mode recognition, thereby achieving higher recognition accuracy and stability.

#### 2.2.1 Attention-Aware Convolutional Neural Network

Traditional CNNs [27], while possessing strong feature extraction capabilities for data with local correlations and structural properties, such as time series and images, often struggle to capture long-range dependencies and global features effectively. To overcome this limitation, attention mechanisms originally popularized by transformer models are designed to focus on the most relevant parts of the data, thereby enhancing the model's ability to extract global features and improve overall prediction performance. In this context, we introduce that the AM-CNN processes diverse input signals from ECE system using a unified dot-product attention mechanism for improved global feature extraction. The convolutional neural network constructed is shown in figure 3, covers the time-sliced processing of 11channels ECE signals from the plasma boundary to the core, storing the processed 1D ECE signal data in a three-dimensional array structure as input data for the input layer. The sampling rate of the original ECE signal is 1 MHz, while the frequency range of tearing mode is approximately 1-20 kHz. Given that the future goal is to apply this model to real-time tear mode recognition, we sought to minimize any preprocessing steps that could negatively impact the model's response time, particularly in the early stages. As a result, we avoided complex data preprocessing techniques that alter the resolution of the raw ECE measurements. This constraint imposed high performance requirements on the model.

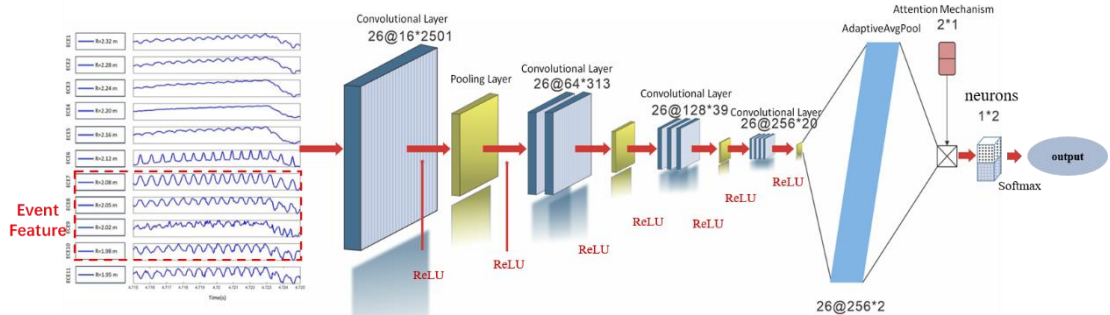

**Figure 3.** The architecture of the segmentation algorithm showing the design of the tearing mode classifier in a multi-channel attention-aware CNN. For a convolutional layer (26@16*2501), each sample generates 16 feature maps, each with a length of 2501, and there are a total of 26 samples or time slices. One sample denotes 11*20002 data and each with 20002 data points, structured as a 3D array 26*11*20002.

This network consists of six layers designed for feature extraction from sequential data, followed by an attention mechanism and a final classification layer. The first four layers (layer 1 to layer 4) each consist of two convolutional blocks with increasing filter sizes (16, 64, 128, 256) and decreasing kernel sizes (16, 8, 4, 2), interspersed with batch normalization, Rectified Linear Unit activations ($f(x)=\max(0, x)$), max pooling, and dropout for regularization. Each convolutional block reduces the feature map size while increasing the depth of features. Afterward, layer 5 applies adaptive average pooling to reduce the output to a fixed size, followed by a fully connected layer in layer 6 that outputs 2 classes using a SoftMax function. A dot-product attention mechanism in Section 2.2.2 is applied on the final output, using a hidden size of 2 (with tearing mode and without tearing mode), to focus on relevant features, enabling the model to flexibly adjust its weight parameters and focus on specific time when key features appear. Finally, connected to the output layer, the output layer consists of two neurons connected to the fully connected layer, activating the output layer neurons through the SoftMax function to generate the probability distribution of the existence of the tearing mode. The overall architecture is designed for sequential data classification, with regularization techniques to prevent overfitting. During model deployment, the training data was standardized to ensure stability, and the model was trained using the Adam optimizer with a cross-entropy loss function. A learning rate scheduler dynamically adjusted the learning rate based on validation loss, while an early stopping mechanism prevented overfitting by halting training if validation loss did not improve after 20 consecutive epochs. The initial learning rate was set to 0.0005, with training capped at 100 epochs and a batch size of 26. To further enhance efficiency, the learning rate was reduced by 90% if the loss failed to improve for 10 epochs, and the model with the best validation performance was saved for future use.

### 2.2.2. Attention mechanism module

The attention mechanism [28] is a specialized structure embedded in machine learning models that automatically learns and calculates the contribution of input data to output data. In deep learning, the attention mechanism helps the model discern the importance of different pieces of local information. In a deep convolutional neural network (CNN) model for image recognition, the local information of the image is typically extracted through convolutional kernels. However, the contribution of each piece of local information to the correct recognition of the image varies. Hence, the introduction of the attention mechanism enables the model to focus more on the local information that significantly impacts the recognition results, thereby enhancing the model's performance. The attention mechanism simplifies the model, accelerates computation, and alleviates the problem of long-distance memory, making the model more effective in processing large-scale data.

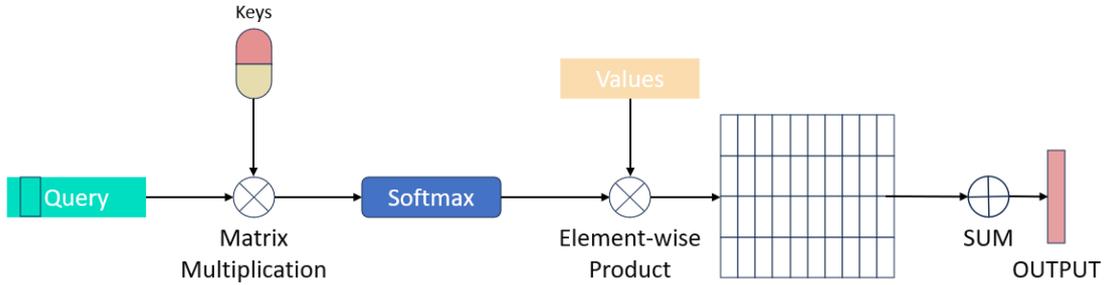

**Figure 4. An illustration of the dot-product attention mechanism, showing the process of computing attention weights using query, key, and value, and generating the weighted output.**

The concept of attention resembles the idea of addressing. For a given query in the target, the similarity or relevance between the query and each key is calculated to determine the weight coefficients for the corresponding values. These weights are then used to compute a weighted sum of the values, resulting in the final attention value. The resulting matrix of outputs is then derived through the following computation:

$$\text{Attention}(Q, K, V) = \text{softmax}(QK^T)V.$$

The attention module used in this work is shown in figure 4, where a 2×1 matrix is set as keys, and the importance of the time point is determined by multiplying with the input query matrix and then using the SoftMax function. The points with high importance will get higher values. In the training of the model in this paper, when there is a flat ECE signal or a phase difference in the ECE signals of multiple channels, the attention module of the model can give higher weights to these time points, that is, they will have higher attention. Finally, the attention weights are dot-multiplied with the input, and the values of the same time point but different channels are summed. At the time points with key features, the values on the neurons will be higher.

## 3. Results

### 3.1 Accuracy, Recall, and F1-Score Analysis of AM-CNN

The tearing mode on tokamaks is characterized by unique reversal features in temperature signals, which pose challenges for traditional CNN in achieving robust generalization across diverse cases. To address this, we compare the performance of a standard CNN with a multi-channel attention-aware convolutional neural network (AM-CNN) in performance metrics which includes three key algorithm evaluation indexes, accuracy, recall and F1-score. This comparison aims to evaluate whether the inclusion of an attention mechanism can enhance the model's ability to extract and classify key features from complex signal patterns. As shown in figure 5, the accuracy of the AM-CNN shows an improvement, reflecting its good ability to correctly identify both tearing mode and non-tearing mode cases. Precision measures the proportion of instances in which the model predicts positive samples to be positive, recall reflects the proportion of real positive samples correctly predicted by the model, and the F1-score is the harmonic average of accuracy and recall, taking into account the performance of both. The recall scores indicate that the AM-CNN is more effective in minimizing false negatives, which is critical in identifying tearing mode events. Similarly, the F1-score, which balances precision and recall, highlights the enhanced overall classification capability of the AM-CNN. This improvement underscores the role of the attention mechanism in focusing on relevant features while mitigating noise from irrelevant data.

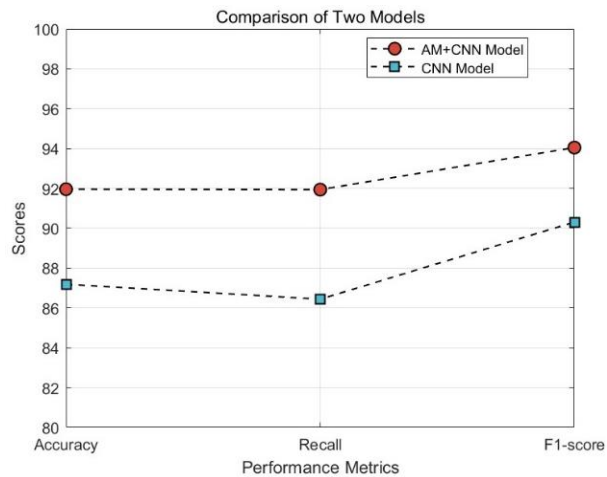

**Figure 5. Performance comparison of the two models CNN and AM-CNN based on algorithmic evaluation metrics including accuracy, recall and F1-score.**

It should be noted that the two models use the same training set and test set to give classification results. By comparing the three evaluation indicators of the results, it can be found that after the introduction of attention mechanism in the convolutional neural network, the accuracy increases by 4.77%, the recall increases by 5.5%, and the F1 increases by 3.74%. This significant improvement across all key metrics demonstrates the effectiveness of integrating the attention mechanism into the convolutional neural network. The enhanced precision indicates a reduction in false positives, while the increased recall reflects better identification of true positives. Together, these improvements contribute to a higher F1-score, showcasing the good overall performance and robustness of the attention-aware CNN model.

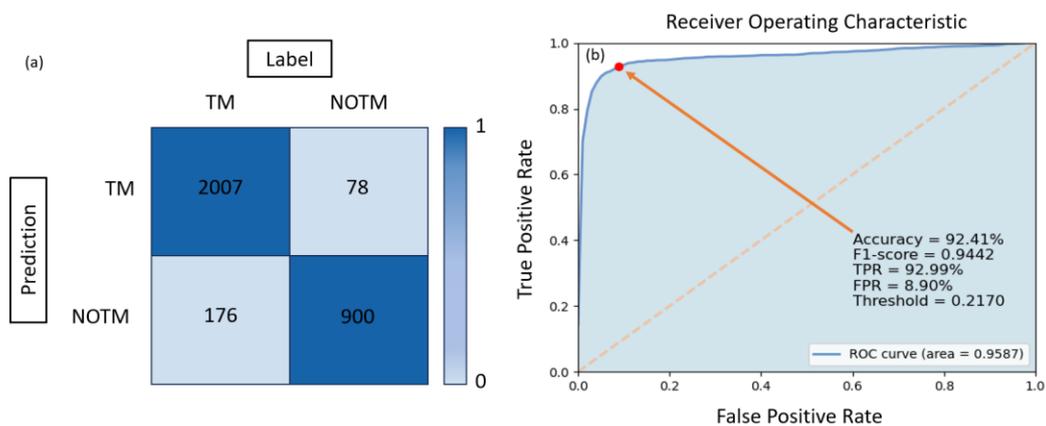

**Figure 6: (a) Confusion matrix of TM classifier on test set. (b) ROC curve of TM classifier on test set.**

Besides, in order to better evaluate the AM-CNN model, the ROC curve and confusion matrix serve as complementary tools to assess the overall performance and the specific strengths and weaknesses. The ROC curve is particularly useful for evaluating the overall classification capability of the model. It helps assess how well the model distinguishes between the two classes (tearing mode and non-tearing mode) by plotting the true positive rate (sensitivity) against the false positive rate at various threshold settings. By analyzing the area under the ROC curve, we can gauge the model's ability to correctly classify both classes across different thresholds, making it an ideal tool for understanding the model's discriminative power and its robustness in various operating conditions. On the other hand, the confusion matrix provides a detailed breakdown of the model's performance on a per-class basis. It shows the number of true positives, true negatives, false positives, and false negatives, which helps identify any biases in the model's predictions.

As shown in figure 6, the performance of the AM-CNN model is further evaluated using the ROC curve and confusion matrix, which provide a more detailed understanding of its classification abilities. The test dataset includes 3161 denoised ECE signals, with 30.94% lacking the tearing mode temperature feature and 69.06% having the tearing mode temperature feature. As shown in Figure 6(a), the confusion matrix provides the accurate counts predicted by the model. Among the 978 samples without tearing patterns, 900 were successfully identified, with a success rate of about 92.02%. For the 2183 samples with tearing mode temperature feature, 2007 are successfully identified, with a success rate of about 91.94%. As shown in Figure 6 (b), the ROC curve shows that the model achieves a high true positive rate (TPR) of 92.99% and a relatively low false positive rate (FPR) of 8.9% at a threshold of 0.2170, indicating that it can effectively distinguish between tearing mode and non-tearing mode events. With a threshold of 0.2170, the model strikes a balance between sensitivity and specificity, reflecting its ability to classify both classes accurately. These results align with the model's high accuracy (92.41%) and F1-score (0.9442), further confirming its strong classification performance, while also highlighting areas for potential improvement, such as reducing false negatives to enhance the detection of rare tearing mode events. Overall, both the ROC curve and confusion matrix demonstrate that the AM-CNN model performs well in identifying tearing modes, with a good balance between precision and recall.

### 3.2 TM classifier for time evolution recognition

The time evolution of tearing mode recognition is analyzed to gain deeper insights into both the physical characteristics of TMs and the performance of the AM-CNN model at different stages of mode development. By visualizing the recognition process over time, we can observe how the model responds to the dynamic features of TMs, from the early seed phase to the formation of a stable magnetic island. Moreover, studying the time evolution helps identify specific performance bottlenecks, such as the model's reduced sensitivity to subtle early-stage features, guiding further optimization for better detection of seed magnetic islands. Such insights not only contribute to improving the model's robustness but also provide valuable guidance for plasma control strategies, where timely identification of TMs is essential to mitigate instabilities in magnetic confinement systems. Thus, the visualization of TM recognition over time serves as an essential tool for both scientific exploration and practical application. This analysis is critical because it highlights the model's challenges during the TM onset phase, where signal features are weak and ambiguous, and its improved performance as the magnetic island grows and the reversal surface becomes more distinct.

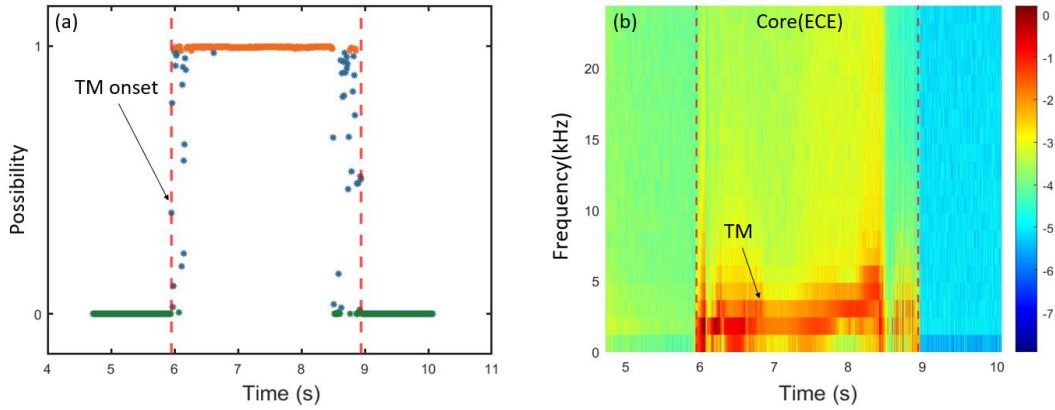

**Figure 7. A Classification results of TM classifier on testing set. (a) Time evolution of the AM-CNN model's prediction results for a randomly selected discharge #86502 event in the test set. (b) Short-time Fourier transform spectrogram of the central ECE signal from the tearing mode discharge event generally including both classical tearing modes and neoclassical tearing modes.**

Figure 7(a) illustrates the evaluation of the AM-CNN model's performance on a randomly selected discharge event from the test set, providing insights into its capability to predict tearing modes (TMs). Through the evolution of the prediction result, it can be observed that the AM-CNN model's prediction probability reaches its stable value during the 7-8 s period. This corresponds to the phase when the tearing mode has developed a stable and sufficiently wide magnetic island, making the reversal surface and associated signal characteristics more distinct and easier for the model to identify. The AM-CNN model is trained 100 times. During the training process, the learning rate is reduced twice in the 81st and 93rd rounds. With the increase in training rounds, the loss function of the training set gradually decreased, and the accuracy continued to improve. Prior to lowering the learning rate, the AM-CNN model showed significant fluctuations in the loss function and accuracy on the test set. In contrast, during the TM onset phase, the prediction probabilities are less stable and exhibit fluctuations. Especially in the time from 5.72 s to 5.96 s at the beginning of the model, the frequency induced by the tearing mode on ECE signal is lower than the frequency during 7-8 s periods. The size and position of the tear mode vary greatly, and the amplitude of the reversal signal generated by the tear mode is also small, but the model can still distinguish the tearability of tearing mode better than the low frequency condition. This reflects the model's difficulty in identifying tearing modes during the early stages of their development, likely due to the limited or ambiguous signal features when the magnetic island is still small and the reversal surface is harder to detect. This challenge underscores the need for further optimization in identifying seed magnetic islands during the onset phase, an area that will require additional attention to enhance the model's sensitivity to subtle early-stage features in future work.

Figure 7(b) shows the short-time Fourier transform (STFT) spectrogram of the central ECE signal for the same discharge #86502. The spectrogram reveals frequency components associated with tearing mode dynamics. The amplitude of temperature fluctuations, indicated by the color intensity, highlights the temporal evolution of the mode, providing a physical basis for the model's prediction behavior. This combined analysis underscores the AM-CNN model's potential for TM classification while identifying specific challenges during the early stages of mode development. As the magnetic island reaches a more stable state, the reversal surface becomes more clearly defined, and the model's predictions improve significantly. The increase in prediction probabilities demonstrates the model's strength in recognizing tearing modes once the characteristic features of the reversal surface are more prominent.

## 4.summary and discussion

In this study, we proposed an attention-aware convolutional neural network (AM-CNN) for identification of tearing modes in tokamak discharges using multi-channel ECE diagnostics from the EAST tokamak. By incorporating an attention mechanism with convolutional layers, our model not only captures the local features of the data but also enhances the extraction of global patterns essential for identifying tearing mode behaviors. The dataset used for model training and validation consisted of $2.5\times10^9$ samples collected from 68 shots spanning from years 2016 to 2021, covering both tearing and non-tearing mode cases. The data were split into training, testing, and validation sets, with the training set constituting 66.5% of the total, and the model demonstrated an impressive classification accuracy of 91.96%, a notable improvement over conventional CNN models. The integration of the attention mechanism enabled the model to focus on the most relevant temporal features in the ECE signals, particularly those linked to the onset and development of magnetic islands, thus improving classification performance in terms of accuracy, recall, and F1 score. The AM-CNN's performance was further evaluated using key metrics, such as the ROC curve and confusion matrix, which highlighted its ability to distinguish tearing mode from non-tearing mode events with high sensitivity (92.99% true positive rate) and low false positive rate (8.9%). This shows that the attention mechanism effectively mitigates noise and focuses the model on the most important features in the tearing mode, thus improving the overall robustness of the model. Moreover, the time evolution of tearing mode recognition demonstrated that the model performs optimally when magnetic islands reach a stable state, with the reversal surface becoming more distinct. However, the model showed challenges in the early stages of tearing mode development when the magnetic islands are small and their signal features are weaker or less distinct. This observation highlights an important area for future improvement, as the model's ability to detect early-stage features is critical for timely intervention in plasma control systems.

The physical understanding of tearing modes with the advanced deep learning framework not only enhances the interpretability and generalization of the model but also lays the groundwork for intelligent control strategies in future ITER. In future work, not only the optimization is needed to enhance sensitivity to early-stage tearing modes, especially during the initial phase when the magnetic island is small and its signal is weak, but also the primary focus will be on identifying the width and position of the tearing mode. This will involve comparing the results of the AM-CNN with those from traditional tearing mode recognition and control algorithms, assessing their accuracy and performance. On the other hand, thorough fine-tuning the parameters of model, it is possible to detect the presence of locked modes on the EAST tokamak in future work. Additionally, to further enhance the predictive capabilities of the model, more diagnostic signals will be incorporated, such as those from magnetic probes and soft X-ray systems. These additional data sources will help explore and identify potential precursors triggered by seed magnetic islands, which could offer earlier detection of instability onset. This approach aims to improve both the accuracy and the time efficiency of tearing mode identification, which is critical for the real-time control of tearing mode in fusion devices like ITER.